\documentclass[preprint,preprintnumbers,amsmath,amssymb,superscriptaddress]{revtex4}
\usepackage{graphicx}
\usepackage{dcolumn}
\usepackage{bm}
\usepackage{color}

\font\scripti=cmmi7
\font\scriptscripti=cmmi5
\def\sib#1{\setbox0 = \hbox{\scripti #1}
  \kern-.02em\copy0\kern-\wd0
  \kern.04em\box0} 
\def\ssib#1{\setbox0 = \hbox{\scriptscripti #1}
  \kern-.02em\copy0\kern-\wd0
  \kern.04em\box0} 
\font\tenib=cmmib10 
\skewchar\tenib='177 \skewchar\tenib='177 \skewchar\tenib='177
\def\pbold#1{\setbox0 = \hbox{$ #1 $}
  \kern-.022em\copy0\kern-\wd0
  \kern.011em\copy0\kern-\wd0
  \kern.011em\copy0\kern-\wd0
  \kern.011em\copy0\kern-\wd0
  \kern.011em\box0} 

\usepackage{graphicx}
\usepackage{dcolumn}
\usepackage{bm}
\usepackage{mathrsfs}

\def\lesssim{\ \raise.3ex\hbox{$<$}\kern-0.8em\lower.7ex\hbox{$\sim$}\ }
\def\gesim{\ \raise.3ex\hbox{$>$}\kern-0.8em\lower.7ex\hbox{$\sim$}\ }

\begin{document}
\preprint{RIKEN-QHP-500}

\title{Three-body crossover from a Cooper triple to bound trimer state in three-component Fermi gases near a triatomic resonance}

\author{Hiroyuki Tajima}
\email[]{hiroyuki.tajima@phys.s.u-tokyo.ac.jp}
\affiliation{
Department of Physics, Graduate School of Science,
The University of Tokyo, Tokyo 113-0033, Japan
}

\author{Shoichiro Tsutsui}
\email[]{shoichiro.tsutsui@riken.jp}
\affiliation{
	Theoretical Research Division, Nishina Center, RIKEN, Wako, Saitama 351-0198, Japan
}

\author{Takahiro M. Doi}
\email[]{takahiro.doi@rcnp.osaka-u.ac.jp}
\affiliation{
	Research Center for Nuclear Physics (RCNP), Osaka University, 567-0047, Japan
}

\author{Kei Iida}
\email[]{iida@kochi-u.ac.jp}
\affiliation{
Department of Mathematics and Physics, Kochi University, 780-8520, Japan
}

\date{\today}
\begin{abstract}
We theoretically investigate ground-state properties of a three-component Fermi gas with pairwise contact interactions between different components near a triatomic resonance where bound trimers are about to appear.
Using variational equations for in-medium two- and three-body cluster states in three dimensions, we elucidate the competition of pair and triple formations due to the Fermi surface effects.
We present the ground-state phase diagram that exhibits transition from a Cooper pair to Cooper triple state and crossover from a Cooper triple to tightly bound trimer state at negative scattering lengths.
This three-body crossover is analogous to the Bardeen-Cooper-Schrieffer to Bose-Einstein condensation crossover observed in a two-component Fermi gas.
We predict that the threshold scattering length $a_{-}$ for three-body states can be shifted towards the weak-coupling side due to the 
 emergence of Cooper triples.
\end{abstract}
\maketitle
\section{Introduction}
A cold atomic system is an ideal platform for quantum simulations of various few- and many-body problems~\cite{BlochRMP,GiorginiRMP}.
In particular, the Feshbach resonance enables us to control 
 the $s$-wave scattering length $a$ characterizing the interaction strength between atoms~\cite{ChinRMP}.
Moreover, thanks to a simplicity of their models, one can systematically compare theoretical results with experimental ones even in the strong-coupling regime.

Experimental realization of the crossover from 
the weak-coupling Bardeen-Cooper-Schrieffer (BCS) Fermi superfluid to the Bose-Einstein-condensation (BEC) of tightly bound dimers with increasing strength of the pairing interaction is one of the most important breakthroughs in this atomic system~\cite{Regal2004,Zwierlein2004,Bartenstein2004,Kinast2004}.
Physical properties of two-component Fermi gases throughout the BCS-BEC crossover have been studied in detail theoretically and experimentally~\cite{Zwerger,Randeria,Strinati,Ohashi}. 
Nowadays, the BCS-BEC crossover has attracted tremendous attention from communities of condensed matter~\cite{Kasahara,Hashimoto,Nakagawa} and nuclear physics~\cite{Sun,Durel,Margueron,Strinati,Ohashi}. 

Another important example of non-trivial physics simulated in cold atoms is the Efimov effect~\cite{Efimov1970}.
An infinite series of three-body bound states with a discrete scale symmetry arises near the unitarity limit ($|a|\rightarrow\infty$)~\cite{Braaden,Greene,NaidonEndo}.
While the Efimov effect was predicted in the context of nuclear physics,
such a non-trivial state was first observed in cold atomic systems via the measurement of three-body losses~\cite{Kraemer,Zaccanti}. 
Recently, the Efimov physics has also been explored in spin systems~\cite{Nishida2013} as well as in helium atoms~\cite{Kunitski}.
It should be noted that various exotic phenomena associated with three-body physics have also been discussed in ultracold atoms~\cite{NishidaS,Deng,NishidaSemi,Deng2,Musolino}.

A three-component Fermi gas, which can be realized in recent experiments, offers an opportunity to investigate a unique interplay between pair and trimer formations.
Indeed, this system has been anticipated as a quantum simulator of color superconductivity in dense quantum chromodynamics (QCD) matter~\cite{Ohara}. 
In this context, the superfluid state has theoretically been explored in a three-component Fermi gas with attractive interactions~\cite{Paananen,He2006,Ozawa,Salasnich}.
In $^6$Li experiments for three-component mixtures, on the other hand,
three-body physics such as the Efimov effect can be observed via three-body loss measurement~\cite{Williams,Huckans,Nakajima2,Wenz,Lompe2,Huang} as well as the radio-frequency photoassociation~\cite{Nakajima,Lompe}.
While the Fermi degeneracy can be achieved experimentally~\cite{Ottenstein},
the superfluid state has not yet been realized due to the strong three-body loss near the Feshbach resonance.
In this regard, numerous attention is paid to understand how the Efimov trimer behaves in the presence of Fermi seas~\cite{MacNeill,Nygaard,Bellotti,Nishida2015,Sun2019,Alhyder,Enss2020,Sanayei20202,Sanayei2020}.
Moreover, a unique crossover phenomenon involving superfluidity and trimer gases has been investigated~\cite{Floerchinger,Nishida2012,Kirk,Tajima20192,Tajima2019}.
In one dimension, the phase diagram has been explored by using the Bethe ansatz~\cite{Liu,He2010,Kuhn}.
It is also an interesting question whether a three-body counterpart of a Cooper pair called a Cooper triple exists or not in three-component Fermi mixtures~\cite{Niemann,Kirk,Tajima2020,Akagami2021}.

In this paper, we investigate ground-state properties of a three-component Fermi gas with two-body attractive interactions near the triatomic resonance $a=a_{-}$ where a bound trimer starts to appear in a three-body system.
Since three-component Fermi gases are dominated by the formation of tightly bound trimers near the magnetic Feshbach resonance, we focus on the relatively weak-coupling regime to discuss the competition between two- and three-body clusters.
While the attractive contact interactions among three atoms 
exhibit infinitely deep three-body bound states known as the Thomas collapse~\cite{Thomas},
we utilize the momentum cutoff (which corresponds to the inverse range of two-body interactions) to avoid the collapse.
Using a variational method, we show that weakly bound trimers assisted by the Fermi surface effect, that is, Cooper triples, continuously change to tightly bound trimers with increasing two-body attraction.
This three-body crossover is analogous to the BCS-BEC crossover in a two-component Fermi gas in the sense that these crossovers occur around three- and two-body resonances, respectively, where the corresponding bound states start to appear.
We propose the ground-state phase diagram in the space of the scattering length and the range parameter.
It is in contrast to the previous works in which the phase diagram is shown in terms of the negative effective range associated with the Feshbach coupling in the two-channel model~\cite{Nishida2012,Tajima2019}. 
Moreover, we give a physical interpretation of the shift of three-body loss due to the Fermi surface effect as an emergence of Cooper triples.
Considering the non-Hermitian three-body interaction responsible for the three-body loss~\cite{Kirk}, we evaluate the medium-induced three-body loss rate within the variational approach. 
It is shown that the three-body loss is enhanced by the Cooper triple formation compared to the case without Fermi seas.

This paper is organized as follows.
In Sec.~\ref{sec2}, we show the Hamiltonian for a three-component Fermi gas with two-body contact interactions.
In Sec.~\ref{sec3}, we present a variational method for in-medium two- and three-body states on top of the Fermi sea.
In Sec.~\ref{sec4}, we show our numerical results for the in-medium bound states, the ground-state phase diagram, and the three-body loss rate.
Finally, we summarize this paper in Sec.~\ref{sec5}.
In what follows, we take $\hbar=k_{\rm B}=1$ and the system volume is taken to be unity.

\section{Hamiltonian}
\label{sec2}
We consider a homogeneous three-component Fermi gas in three dimensions, which is described by the Hamiltonian
\begin{align}
\hat{H}=\hat{K}+\hat{V}_{12}+\hat{V}_{23}+\hat{V}_{13}
\end{align}
with the kinetic term
\begin{align}
\hat{K}=\sum_{\gamma=1,2,3}\sum_{\bm{p}}\xi_{\bm{p},\gamma}\hat{c}_{\bm{p},\gamma}^\dag \hat{c}_{\bm{p},\gamma}
\end{align}
and the contact-type intercomponent interaction term
\begin{align}
\hat{V}_{\gamma\gamma'}=g_{\gamma\gamma'}\sum_{\bm{k}_1,\bm{k}_2,\bm{k}_1',\bm{k}_2'}\hat{B}_{\bm{k}_1\gamma,\bm{k}_2\gamma'}^\dag
\hat{B}_{\bm{k}_1'\gamma,\bm{k}_2'\gamma'}\delta_{\bm{k}_1+\bm{k}_2,\bm{k}_1'+\bm{k}_2'},
\end{align}
where $\gamma=1,2,3$ denotes the hyperfine states which we call ``color" in analogy with QCD (note that $\gamma\neq\gamma'$).
$\xi_{\bm{p},\gamma}=p^2/(2m_\gamma)-\mu_\gamma$ is the kinetic energy of a Fermi atom with mass $m_\gamma$ and momentum $\bm{p}$, measured from the chemical potential $\mu_\gamma$, while
$\hat{c}_{\bm{p},\gamma}$ and $\hat{c}_{\bm{p},\gamma}^\dag$ are the corresponding annihilation and creation operators.
For convenience, we have defined the pair creation and annihilation operators as
\begin{align}
    \hat{B}_{\bm{k}_1\gamma,\bm{k}_2\gamma'}^\dag=\hat{c}_{\bm{k}_1,\gamma}^\dag \hat{c}_{\bm{k}_2,\gamma'}^\dag,
\end{align}
\begin{align}
    \hat{B}_{\bm{k}_1\gamma,\bm{k}_2\gamma'}=\hat{c}_{\bm{k}_2,\gamma'} \hat{c}_{\bm{k}_1,\gamma}. 
\end{align}
For simplicity, we consider the case of equal masses $m_1=m_2=m_3\equiv m$, equal chemical potentials $\mu_1=\mu_2=\mu_3\equiv \mu$, and equal coupling constants $g_{12}=g_{23}=g_{13}\equiv g$.
Simultaneously we define $\xi_{\bm{p}}=p^2/(2m)-\mu$.
The coupling constant $g$ is related to the $s$-wave scattering length $a$ as
\begin{align}
    \frac{m}{4\pi a}=\frac{1}{g}+\frac{m\Lambda}{2\pi^2},
\end{align}
where $\Lambda$ is the momentum cutoff which is kept finite to avoid the Thomas collapse~\cite{Thomas}.
Also, we do not consider the three-body interaction whose effect has already been investigated in Refs.~\cite{Drut,McKenny,Tajima2020,Akagami2021}.
While the three-body attraction tends to stabilize three-body bound states, here we are interested in the stabilization of them due to the purely two-body attractions.

\section{Variational ansatz}
\label{sec3}
We are now in a position to calculate the energies of a single two-body and a single three-body state above the Fermi sea.
\subsection{Variational wave function for a Cooper pair}
First, we consider the variational wave function for a Cooper pair consisting of $\gamma$ and $\gamma'$ components given by
\begin{align}
    |\Psi_{\rm CP}\rangle = \sum_{|\bm{p}|\geq k_{\rm F}}\Phi_{\bm{p}}\hat{B}_{\bm{p}\gamma,-\bm{p}\gamma'}^\dag |{\rm FS}\rangle,
\end{align}
where $\Phi_{\bm{p}}$ is the variational parameter.
Below, we consider an $s$-wave Cooper pair, and assume that $\Phi_{\bm{p}}=\Phi_{|\bm{p}|}$.
The Fermi sphere $|{\rm FS}\rangle$ is defined by
\begin{align}
|{\rm FS}\rangle=\prod_{\gamma}\prod_{|\bm{k}|\leq k_{\rm F}}\hat{c}_{\bm{k},\gamma}^\dag |0\rangle    
\end{align}
where $k_{\rm F}$ is the Fermi momentum ($|0\rangle$ is a vacuum state).
Although we have three possibilities for the choice of $(\gamma,\gamma')$, i.e., (1,2), (2,3), and (1,3), these three Cooper pairs on top of the Fermi sea are degenerate under the U(3) invariance of interactions.
In a BCS pairing state, such a symmetry can be broken spontaneously in such a way as to lead to population imbalance~\cite{Ozawa,Salasnich}; we do not go into detail about this possibility since it is out of scope in this paper.   
\par
Using $|\Psi_{\rm CP}\rangle$, one can evaluate the ground-state energy $E_{\rm GS}=\langle \Psi_{\rm CP}| \hat{H} |\Psi_{\rm CP}\rangle$ as
\begin{align}
   E_{\rm GS}=\langle \Psi_{\rm CP}|\hat{K}|\Psi_{\rm CP} \rangle + \langle \Psi_{\rm CP}| \hat{V}_{\gamma\gamma'}|\Psi_{\rm CP} \rangle + {\rm const.}, 
\end{align}
where the expectation values of the kinetic term and the $\gamma$--$\gamma'$ interaction term are given by
\begin{align}
    \langle \Psi_{\rm CP}|\hat{K}|\Psi_{\rm CP} \rangle=\sum_{|\bm{p}|\geq k_{\rm F}}|\Phi_{\bm{p}}|^2\left(\xi_{\bm{p}}+\xi_{-\bm{p}}\right)
\end{align}
and
\begin{align}
    \langle \Psi_{\rm CP}| \hat{V}_{\gamma\gamma'}|\Psi_{\rm CP} \rangle=g\sum_{|\bm{p}|\geq k_{\rm F}}\sum_{|\bm{p}'|\geq k_{\rm F}}\Phi_{\bm{p}}^*\Phi_{\bm{p}'},
\end{align}
respectively.
Here we neglected the constant terms which are not relevant for the variational equation. 
Indeed, such terms produce a shift of $\mu$ in $\xi_{\bm{p}}=p^2/(2m)-\mu$.
In what follows, we take $\xi_{\bm{p}}=p^2/(2m)-E_{\rm F}$,
where the Fermi energy $E_{\rm F}=k_{\rm F}^2/(2m)$ may be assumed to include such irrelevant shifts.
Using the Lagrange multiplier $E_2$ associated with the constraint on the norm given by $\langle \Psi_{\rm CP}|\Psi_{\rm CP}\rangle=1$, we obtain the variational equation
\begin{align}
    \frac{\delta}{\delta \Phi_{\bm{p}}^*}\Bigl(E_{\rm GS}-E_2\langle \Psi_{\rm CP}|\Psi_{\rm CP}\rangle\Bigr)=0.
\end{align}
The explicit form of the equation for $\Phi_{\bm{p}}$ ($|\bm{p}|\geq k_{\rm F}$) reads
\begin{align}
    (2\xi_{\bm{p}}-E_2)\Phi_{\bm{p}}+g\sum_{|\bm{p}'|\geq k_{\rm F}}\Phi_{\bm{p}'}=0,
\end{align}
which can be rewritten as the equation for $E_2$ given by 
\begin{align}
\label{eqCP}
    1=-g\sum_{|\bm{p}|\geq k_{\rm F}}\frac{1}{p^2/m-2E_{\rm F}-E_2}.
\end{align}
Indeed, Eq.~(\ref{eqCP}) is the well-known equation for the Cooper problem and consistent with the diagrammatic approach~\cite{Niemann} where the Fermi surface effects are treated as the lower boundary of the momentum integral.
From the comparison of the pole of the two-body $T$-matrix
\begin{align}
    T_2(\omega_2)=g\left[1+g\sum_{\bm{p}}\frac{1}{p^2/m-\omega_2-i\delta}\right]^{-1}
\end{align}
(where $\omega_2$ is the two-body energy and $\delta$ is a positive infinitesimal) with Eq.~(\ref{eqCP}), the in-medium two-body pole can be regarded as $E_{\rm pole}^{\rm 2b}=2E_{\rm F}+E_2$.
Eventually, $E_2<0$ corresponds to the binding energy of a Cooper pair.
Since Eq.~(\ref{eqCP}) reproduces the binding energies of a weak-coupling Cooper pair and a tightly bound diatomic molecule in both limits, $E_{\rm pole}^{\rm 2b}/2$ shows a similar behavior of the solution of $\mu$ within the BCS-Leggett theory throughout the BCS-BEC crossover~\cite{Niemann}. 

\subsection{Variational wave function for a Cooper triple}
Next, we consider the variational wave function for a Cooper triple given by
\begin{align}
\label{eqCT1}
    |\Psi_{\rm CT}\rangle = \sum_{|\bm{k}_1|\geq k_{\rm F}}\sum_{|\bm{k}_2|\geq k_{\rm F}}\sum_{|\bm{k}_3|\geq k_{\rm F}}\mathcal{O}_{\bm{k}_1,\bm{k}_2,\bm{k}_3}\hat{C}_{\bm{k}_1,\bm{k}_2,\bm{k}_3}^\dag\delta_{\bm{k}_1+\bm{k}_2+\bm{k}_3,\bm{0}}|{\rm FS}\rangle,
\end{align}
where
\begin{align}
    \hat{C}_{\bm{k}_1,\bm{k}_2,\bm{k}_3}^\dag=\frac{1}{6}\sum_{\gamma_1,\gamma_2,\gamma_3}\varepsilon_{\gamma_1\gamma_2\gamma_3}\hat{c}_{\bm{k}_1,\gamma_1}^\dag \hat{c}_{\bm{k}_2,\gamma_2}^\dag \hat{c}_{\bm{k}_3,\gamma_3}^\dag
\end{align}
is the creation operator of a color-singlet trimer.
We assume that the variational parameter $\mathcal{O}_{\bm{k}_1,\bm{k}_2,\bm{k}_3}$ is a symmetric tensor.
Using the symmetry of $\mathcal{O}_{\bm{k}_1,\bm{k}_2,\bm{k}_3}$,
we can rewrite Eq.~(\ref{eqCT1}) as
\begin{align}
\label{eqCTw}
    |\Psi_{\rm CT}\rangle = \sum_{|\bm{k}_1|\geq k_{\rm F}}\sum_{|\bm{k}_2|\geq k_{\rm F}}\Omega_{\bm{k}_1,\bm{k}_2}\hat{F}_{\bm{k}_1,\bm{k}_2}^\dag|{\rm FS}\rangle,
\end{align}
where
\begin{align}
\label{eqF}
    \hat{F}_{\bm{k}_1,\bm{k}_2}^\dag = \hat{c}_{\bm{k}_1,1}^\dag \hat{c}_{\bm{k}_2,2}^\dag \hat{c}_{-\bm{k}_1-\bm{k}_2,3}^\dag
\end{align}
and $\mathcal{O}_{\bm{k}_1,\bm{k}_2,-\bm{k}_1-\bm{k}_2}=\Omega_{\bm{k}_1,\bm{k}_2}$ without loss of generality.
Since we consider three fermions on top of the Fermi sea,
$\Omega_{\bm{k}_1,\bm{k}_2}$ involves an additional constraint which requires $\Omega_{\bm{k}_1,\bm{k}_2}=0$ for $|-\bm{k}_1-\bm{k}_2|< k_{\rm F}$.
Using this, we can calculate the ground-state energy $E_{\rm GS}=\langle\Psi_{\rm CT}|\hat{H}|\Psi_{\rm CT}\rangle$ as
\begin{align}
    \langle\Psi_{\rm CT}|\hat{H}|\Psi_{\rm CT}\rangle
    &=\langle\Psi_{\rm CT}|\hat{K}|\Psi_{\rm CT}\rangle
    +\langle\Psi_{\rm CT}|\hat{V}_{12}|\Psi_{\rm CT}\rangle
    +\langle\Psi_{\rm CT}|\hat{V}_{23}|\Psi_{\rm CT}\rangle\cr
    &+\langle\Psi_{\rm CT}|\hat{V}_{13}|\Psi_{\rm CT}\rangle +{\rm const}.
\end{align}
The expectation value of the kinetic term reads
\begin{align}
    \langle\Psi_{\rm CT}|\hat{K}|\Psi_{\rm CT}\rangle=\sum_{|\bm{k}_1|\geq k_{\rm F}}\sum_{|\bm{k}_2|\geq k_{\rm F}}|\Omega_{\bm{k}_1,\bm{k}_2}|^2\Bigl(\xi_{\bm{k}_1}+\xi_{\bm{k}_2}+\xi_{-\bm{k}_1-\bm{k}_2}\Bigr)+{\rm const.},
\end{align}
where we omitted the irrelevant terms as we mentioned in the previous section for the wave function of a Cooper pair.
The expectation values of the interaction terms can also be obtained as
\begin{align}
    \langle\Psi_{\rm CT}|\hat{V}_{12}|\Psi_{\rm CT}\rangle&=g\sum_{|\bm{k}_1|\geq k_{\rm F}}\sum_{|\bm{k}_2|\geq k_{\rm F}}\sum_{|\bm{k}_2'|\geq k_{\rm F}}
    \Omega_{\bm{k}_1,\bm{k}_2}^*\Omega_{\bm{k}_1+\bm{k}_2-\bm{k}_2',\bm{k}_2'},\\
\langle\Psi_{\rm CT}|\hat{V}_{23}|\Psi_{\rm CT}\rangle&=g\sum_{|\bm{k}_1|\geq k_{\rm F}}\sum_{|\bm{k}_2|\geq k_{\rm F}}\sum_{|\bm{k}_2'|\geq k_{\rm F}}
    \Omega_{\bm{k}_1,\bm{k}_2}^*\Omega_{\bm{k}_1,\bm{k}_2'},\\
    \langle\Psi_{\rm CT}|\hat{V}_{13}|\Psi_{\rm CT}\rangle&=g\sum_{|\bm{k}_1|\geq k_{\rm F}}\sum_{|\bm{k}_2|\geq k_{\rm F}}\sum_{|\bm{k}_2'|\geq k_{\rm F}}
    \Omega_{\bm{k}_1,\bm{k}_2}^*\Omega_{\bm{k}_1',\bm{k}_2}.
\end{align}
The variational equation 
\begin{align}
    \frac{\delta}{\delta\Omega_{\bm{k}_1,\bm{k}_2}^*}\Bigl(E_{\rm GS}-E_3\langle\Psi_{\rm CT}|\Psi_{\rm CT}\rangle\Bigr)=0
\end{align}
leads to an explicit equation for $\Omega_{\bm{k}_1,\bm{k}_2}$ as
\begin{align}
    \Omega_{\bm{k}_1,\bm{k}_2}(\xi_{\bm{k}_1}+\xi_{\bm{k}_2}+\xi_{-\bm{k}_1-\bm{k}_2}-E_3)=-g\sum_{|\bm{K}|\geq k_{\rm F}}(\Omega_{\bm{k}_1+\bm{k}_2-\bm{K},\bm{K}}+\Omega_{\bm{k}_1,\bm{K}}+\Omega_{\bm{K},\bm{k}_2}).
\end{align}
Here, we define 
\begin{align}
    \mathcal{A}_{1}(\bm{k}_1)=\sum_{|\bm{K}|\geq k_{\rm F}}\Omega_{\bm{k}_1,\bm{K}},
\end{align}
\begin{align}
    \mathcal{A}_{2}(\bm{k}_2)=\sum_{|\bm{K}|\geq k_{\rm F}}\Omega_{\bm{K},\bm{k}_2},
\end{align}
\begin{align}
    \mathcal{A}_{3}(\bm{k}_3\equiv-\bm{k}_1-\bm{k}_2)=\sum_{|\bm{K}|\geq k_{\rm F}}\Omega_{-\bm{k}_3-\bm{K},\bm{K}}.
\end{align}
Using them, we obtain
\begin{align}
    \Omega_{\bm{k}_1,\bm{k}_2}=-g\frac{
    \mathcal{A}_{1}(\bm{k}_1)+A_2(\bm{k}_2)+\mathcal{A}_3(-\bm{k}_1-\bm{k}_2)}{\xi_{\bm{k}_1}+\xi_{\bm{k}_2}+\xi_{-\bm{k}_1-\bm{k}_2}-E_3}.
\end{align}
Taking the momentum summation of $\Omega_{\bm{k}_1,\bm{k}_2}$, we obtain the equations of $\mathcal{A}_{i}(\bm{p})$ as
\begin{align}
\label{eqCT0}
    \mathcal{A}_{i}(\bm{p})\left[
    \frac{1}{g}+\sum_{|\bm{K}|\geq k_{\rm F}}\frac{\theta(\xi_{\bm{p}+\bm{K}})}{\xi_{-\bm{p}-\bm{K}}+\xi_{\bm{K}}+\xi_{\bm{p}}-E_3}\right]
    &=-\sum_{|\bm{K}|\geq k_{\rm F}}
    \frac{\theta(\xi_{\bm{p}+\bm{K}})[
    \mathcal{A}_{k}(-\bm{p}-\bm{K})+\mathcal{A}_j(\bm{K})]}{\xi_{-\bm{p}-\bm{K}}+\xi_{\bm{K}}+\xi_{\bm{p}}-E_3},
\end{align}
for $i\neq j\neq k$.
In Eq.~(\ref{eqCT0}), $\theta(\xi_{\bm{p}+\bm{K}})$ originates from the constraint on $\Omega_{\bm{p},\bm{K}}$ where $\Omega_{\bm{p},\bm{K}}=0$ at $|\bm{p}+\bm{K}|<k_{\rm F}$.
In our symmetric model, we can take $\mathcal{A}_1(\bm{p})=\mathcal{A}_2(\bm{p})=\mathcal{A}_3(\bm{p})\equiv \mathcal{A}(\bm{p})$,
leading to
\begin{align}
\label{eqCT}
    \mathcal{A}(\bm{p})\left[\frac{1}{g}+\sum_{|\bm{K}|\geq k_{\rm F}}\frac{\theta(\xi_{\bm{p}+\bm{K}})}{\xi_{\bm{p}+\bm{K}}+\xi_{\bm{K}}+\xi_{\bm{p}}-E_3}
    \right]
    &=-2\sum_{|\bm{K}|\geq k_{\rm F}}\frac{\theta(\xi_{\bm{p}+\bm{K}})
    \mathcal{A}(\bm{K})}{\xi_{\bm{p}+\bm{K}}+\xi_{\bm{K}}+\xi_{\bm{p}}-E_3}.
\end{align}
We note that a similar in-medium three-body equation can be derived in two dimensions~\cite{Kirk} by summing three equations for $A_{i}(\bm{p})$ over $i$.
Indeed, Eq.~(\ref{eqCT}) is equivalent to the Skorniakov-Ter-Martirosian (STM) equation~\cite{STM} for a three-body problem in vacuum when we set $k_{\rm F}\rightarrow 0$.  In this limit, Eq.\ (\ref{eqCT}) reduces to
\begin{align}
\label{eqSTM}
    \mathcal{A}_0(\bm{p})\left[\frac{1}{g}+\sum_{\bm{K}}\frac{m}{K^2+p^2+\bm{K}\cdot\bm{p}-m\omega_3}
    \right]
    &=-2\sum_{\bm{K}}\frac{
    m\mathcal{A}_0(\bm{K})}{K^2+p^2+\bm{K}\cdot\bm{p}-m\omega_3},
\end{align}
where $\mathcal{A}_0(\bm{p})$ and $\omega_3$ are the three-body amplitude and energy in a three-body problem, and $p=|\bm{p}|$ as well as $K=|\bm{K}|$. In performing the momentum integrals in Eqs.~(\ref{eqCT}) and (\ref{eqSTM}), we use the momentum cutoff $\Lambda$.
Note that the coefficient of $A_0(\bm{p})$ in the left hand side of Eq.~(\ref{eqSTM}) can be rewritten in terms of the inverse two-body $T$-matrix by shifting $\bm{K}\rightarrow \bm{K}-\bm{p}/2$ as
\begin{align}
\frac{1}{g}+\sum_{\bm{K}}\frac{1}{\xi_{\bm{K}+\bm{p}/2}+\xi_{-\bm{K}+\bm{p}/2}+\xi_{\bm{p}}-\omega_3}\equiv T_2^{-1}\left(\omega_3-\frac{3p^2}{4m}\right).
\end{align}
$T_2^{-1}(\omega_2)$ for $\omega_2<0$ can analytically be obtained as
\begin{align}
    T_2^{-1}(\omega_2)&=\frac{m}{4\pi a}-\frac{m}{2\pi^2}\sqrt{m|\omega_2|}\tan^{-1}\left(\frac{\Lambda}{\sqrt{m|\omega_2|}}\right)\cr
    &\rightarrow \frac{m}{4\pi a}-\frac{m}{4\pi}\sqrt{m|\omega_2|} \quad (\Lambda\rightarrow \infty).
\end{align}
While the large $\Lambda$ limit of $T_2^{-1}\left(\omega_3-\frac{3p^2}{4m}\right)$ is employed for the study of the Efimov physics within the present zero-range model~\cite{NaidonEndo}, we keep the cutoff finite in the two-body sector since we are interested in the non-universal regime in the sense that the finite cutoff effect is important.
Such a choice modifies the three-body parameter $\kappa_*$ which is defined by the lowest-energy solution of Eq.~(\ref{eqSTM}) as $\omega_3=\frac{\kappa_*^2}{m}$ at unitarity.
In fact, using the large $\Lambda$ limit of $T_2^{-1}\left(\omega_2-\frac{3p^2}{4m}\right)$, we could numerically reproduce $\kappa^* a_{-}\simeq-1.5$ with $\Lambda\simeq 0.18\kappa^*$ obtained in the previous work~\cite{NaidonEndo}.
However, we emphasize that our results are qualitatively unchanged by this treatment. 
\par
In what follows, we explain how to solve Eq.~(\ref{eqCT}) numerically in this work.
To quantify the cutoff effect, we introduce the range parameter
\begin{align}
    r_{\Lambda}=\frac{4}{\pi\Lambda}.
\end{align}
For the two-body Cooper problem characterized by Eq.~(\ref{eqCP}), $r_\Lambda$ is equivalent to the effective range $r_{\rm eff}$~\cite{Ohashi}.
Strictly speaking, however, we note that the meaning of $r_\Lambda$ in Eq.~(\ref{eqCT}) for Cooper triples is slightly different from $r_{\rm eff}$ since the two-body scattering in Eq.~(\ref{eqCT}) is not described in the zero center-of-mass momentum frame.
Nevertheless, such a difference does not make qualitative change in our results.
Indeed, we obtain $\kappa^* r_\Lambda\simeq 0.44$ for the three-body case, which is close to $\kappa^* r_{\rm eff}=0.44\sim 0.52$~\cite{Naidon2014}.
\par
In this work, we focus on the trimer state with zero angular momentum where $\mathcal{A}(\bm{K})$ is a function of $K=|\bm{K}|$.  By setting $\bm{p}\cdot\bm{K}=pK\cos\theta$, we can rewrite Eq.~(\ref{eqCT}) as
\begin{align}
\label{eqI1I2}
    \mathcal{A}(p)\left[\frac{1}{g}+I_1(p,E_{\rm pole})\right]=-2I_2(p,E_{\rm pole}),
\end{align}
where
\begin{align}
\label{eqi1}
    I_1(p,E_{\rm pole})=\sum_{|\bm{K}|\geq k_{\rm F}}\frac{m\theta(|\bm{p}+\bm{K}|-k_{\rm F})}{K^2+p^2+pK\cos\theta-mE_{\rm pole}},
\end{align}
\begin{align}
\label{eqi2}
    I_2(p,E_{\rm pole})=\sum_{|\bm{K}|\geq k_{\rm F}}\frac{m\theta(|\bm{p}+\bm{K}|-k_{\rm F})\mathcal{A}(K)}{K^2+p^2+pK\cos\theta-mE_{\rm pole}}.
\end{align}
Here we have introduced three-body energy pole $E_{\rm pole}=3E_{\rm F}+E_3$ by analogy with the diagrammatic approach~\cite{Niemann}.
The momentum summation in Eqs.~(\ref{eqi1}) and (\ref{eqi2}) can be replaced by the integral over $k_{\rm F}\leq K \leq\Lambda$ as
\begin{align}
    \sum_{|\bm{K}|\geq k_{\rm F}} \theta(|\bm{p}+\bm{K}|-k_{\rm F})\eta(\bm{K})
    &=\frac{1}{4\pi^2}\int_{k_{\rm F}}^{\Lambda}K^2dK
    \int_{-1}^{1}ds \theta(\sqrt{p^2+K^2-2pKs}-k_{\rm F})\eta(\bm{K})
\end{align}
for an arbitrary function $\eta(\bm{K})$, where we take $s=-\cos\theta$.
One can find that the angular integral is bounded by the step function as
\begin{align}
    s\leq \frac{p^2+K^2-k_{\rm F}^2}{2pK}
\end{align}
when $p^2+K^2-k_{\rm F}^2\leq 2pK$.  
In this regard, we introduce 
\begin{align}
\alpha=\theta(|p-K|-k_{\rm F}^2)+\frac{p^2+K^2-k_{\rm F}^2}{2pK}\theta(k_{\rm F}-|p-K|).    
\end{align}
Using $\alpha$, we obtain
\begin{align}
\label{eqI1}
    I_1(p,E_{\rm pole})=\frac{m}{4\pi^2 p}\int_{k_{\rm F}}^{\Lambda}KdK\ln\left(\frac{K^2+p^2+pk -mE_{\rm pole}}{K^2+p^2-pk\alpha -mE_{\rm pole}}\right),
\end{align}
\begin{align}
\label{eqI2}
    I_2(p,E_{\rm pole})=\frac{m}{4\pi^2 p}\int_{k_{\rm F}}^{\Lambda}KdK\mathcal{A}(K)\ln\left(\frac{K^2+p^2+pk -mE_{\rm pole}}{K^2+p^2-pk\alpha -mE_{\rm pole}}\right).
\end{align}
In the practical calculation, we solve Eq.~(\ref{eqCT}) with an iteration method numerically evaluating Eqs.~(\ref{eqI1}) and (\ref{eqI2}). 
Although $\mathcal{A}(\bm{p})$ has a physical dimension,
there is a scale invariance of Eq.~(\ref{eqI1I2}) via $\mathcal{A}(p)\rightarrow\lambda\mathcal{A}(p)$ for an arbitrary number $\lambda$.
Thus we start the iteration from the initial value $\mathcal{A}(p)=1$ for given value of $E_3$ since $E_3$ is unchanged by the scale transformation of the solution $\mathcal{A}(p)$ in Eq.~(\ref{eqI1I2}).
For the convergence of $\mathcal{A}(p)$,
we have required 
\begin{align}
    \sum_{n}[\mathcal{A}_{\rm in}(n)-\mathcal{A}_{\rm out}(n)]^2\leq 10^{-8},
\end{align}
where $\mathcal{A}_{\rm in/out}(n)$ is the input/output for Eq.~(\ref{eqI1I2}) at the $n$-th discretized momentum $p=n\Delta p+k_{\rm F}$ used in the Newton-Cotes integration with $\Delta p=(\Lambda-k_{\rm F})/N$ (i.e., $\mathcal{A}_{\rm out}(p)=-2I_2(p,E_{\rm pole})/[g^{-1}+I_{1}(p,E_{\rm pole})]$, where $\mathcal{A}_{\rm in}(n)$ is substituted to $I_{2}(p,E_{\rm pole})$). 
We have confirmed that $N=1000$ is sufficient for the convergence in the regime of interest here.

\section{Results}
\label{sec4}
\begin{figure}[t]
    \centering
    \includegraphics[width=8.5cm]{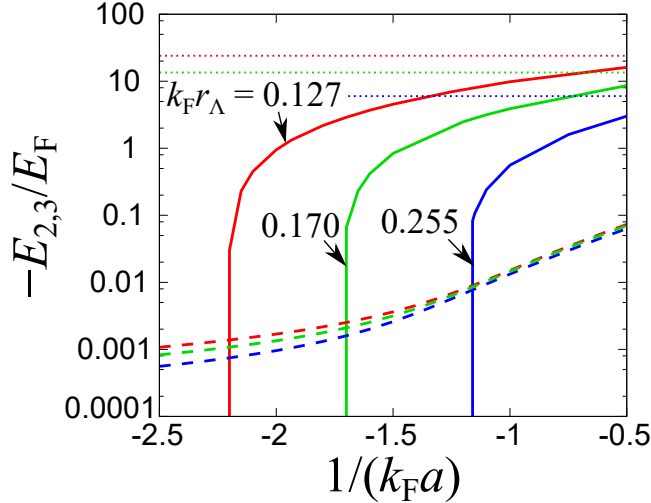}
    \caption{In-medium binding energies $-E_3$ (solid curve) and $-E_2$ (dashed curve) of a trimer and a pair, respectively, calculated at $k_{\rm F}r_{\Lambda}=0.127$, $0.170$, and $0.255$ (corresponding to $\Lambda/k_{\rm F}=10$, $7.5$, and $5$).
    For reference, we plot a cutoff energy scale $0.12\Lambda^2/m$ (horizontal dotted lines) corresponding to the in-vacuum trimer binding energy with each cutoff at unitarity.
    }
    \label{fig:1}
\end{figure}
We proceed to show the results from the variational calculations mentioned above. First, we discuss the condition for the formation of a Cooper pair and a Cooper triple.
Figure~\ref{fig:1} shows the binding energies of an in-medium trimer $-E_3$ and pair $-E_2$ obtained from Eqs.~(\ref{eqCP}) and (\ref{eqCT}), respectively, as functions of the inverse two-body scattering length $1/(k_{\rm F}a)$ at several range parameters $k_{\rm F}r_\Lambda=0.127$, $0.170$, and $0.255$. Although there is no in-vacuum two-body bound state at negative scattering length, $-E_2$ is always finite even in the weak-coupling limit due to the Cooper instability.
On the other hand, $-E_3$ becomes finite above a nonzero critical scattering length.
Once $-E_3$ becomes nonzero, it rapidly increases with increasing $1/(k_{\rm F}a)$ and eventually becomes of the order of $\Lambda^2/m$, 
reflecting the consequence of the Thomas collapse~\cite{Thomas}.
Indeed, a similar behavior was reported in Ref.~\cite{Niemann}.

\begin{figure}[t]
    \centering
    \includegraphics[width=8.5cm]{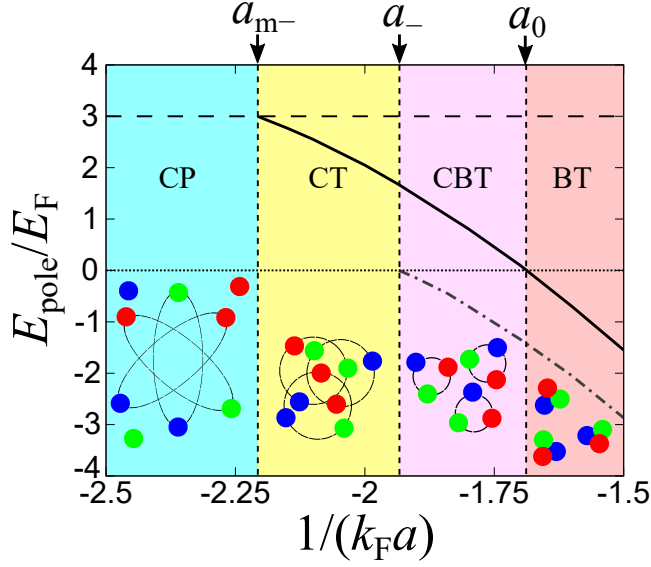}
    \caption{Three-body pole $E_{\rm pole}=3E_{\rm F}+E_3$ (solid curve) for $k_{\rm F}r_{\Lambda}=0.127$. For comparison, we plot the three-body energy $E_{3,{\rm v}}$ (dashed-dotted curve) in vacuum.
    We find four regimes consisting of Cooper pairs (CP, $1/(k_{\rm F}a)<1/(k_{\rm F}a_{{\rm m}-})$), Cooper triples (CT, $1/(k_{\rm F}a_{{\rm m}-})\leq 1/(k_{\rm F}a)<1/(k_{\rm F}a_{{\rm m}-})$), correlated bound trimers (CBT, $1/(k_{\rm F}a_{-})\leq 1/(k_{\rm F}a)<1/(k_{\rm F}a_0)$), and bound trimers (BT, $1/(k_{\rm F}a)\geq 1/(k_{\rm F}a_0)$).  
    }
    \label{fig:2}
\end{figure}
By the critical condition for $E_3<0$ as can be seen from Fig.~\ref{fig:1}, we define the threshold scattering length $a_{{\rm m}-}$ for the in-medium trimer formation. 
In general, $a_{{\rm m}-}$ is different from the vacuum counterpart $a_{-}$.
Figure~\ref{fig:2} shows the three-body pole $E_{\rm pole}=3E_{\rm F}+E_3$ at $k_{\rm F}r_\Lambda=0.127$.
Since there is no three-body bound state in medium at $1/(k_{\rm F}a)<1/(k_{\rm F}a_{{\rm m}-})$, the Cooper pairing (CP) is favored in such a weak-coupling regime.
The transition from the pair state to the trimer state is occurred at $a=a_{{\rm m}-}$.
Note that indeed such a transition may occur at $E_2=E_3$ as shown in Fig.~\ref{fig:1} but it is close to $a=a_{{\rm m}-}$ because of the rapid growth of $-E_3$.
A similar Lifshitz transition from the pair to the trimer state is predicted near $a=a_{-}$ in the two-channel model in Ref.~\cite{Nishida2012} where the medium effect on the trimer state is neglected.
In Fig.~\ref{fig:2}, we also plot the three-body energy $E_{3,{\rm v}}$ in vacuum, which becomes finite at $a=a_{-}$.
This result indicates that the in-medium three-body bound state assisted by the Fermi surface effect exists even in the absence of the in-vacuum counterpart, i.e., in the region of $1/(k_{\rm F}a_{{\rm m}-})<1/(k_{\rm F}a)<1/(k_{\rm F}a_{-})$.
Therefore, we refer to this region as the Cooper triple (CT) state which is indeed the three-body version of Cooper pairing.
At strong coupling where $1/(k_{\rm F}a)\geq 1/(k_{\rm F}a_{-})$, $E_{\rm pole}$ is still positive up to $a=a_0$.
In such a regime, while the in-medium trimer is distinct from the in-vacuum counterpart in 
the sign of $E_{\rm pole}$, the existence of the in-medium trimer is strongly supported by the underlying three-body bound state.
In this sense, we call the regime of $1/(k_{\rm F}a_{-})<1/(k_{\rm F}a)<1/(k_{\rm F}a_0)$ as
the correlated bound trimer (CBT) regime.
Finally, we obtain the negative $E_{\rm pole}$ in a regime of stronger coupling, i.e., $1/(k_{\rm F}a)>1/(k_{\rm F}a_0)$.
In this case, the Fermi surface does not play a significant role in the trimer formation, which is in fact dominated by three-body physics. 
Therefore we call the regime of $1/(k_{\rm F}a)>1/(k_{\rm F}a_0)$ as the bound trimer (BT) regime.
Although the latter two boundaries correspond to the crossover,
they are useful to understand the gradual change in the state of the system qualitatively.

\begin{figure}[t]
    \centering
    \includegraphics[width=8.5cm]{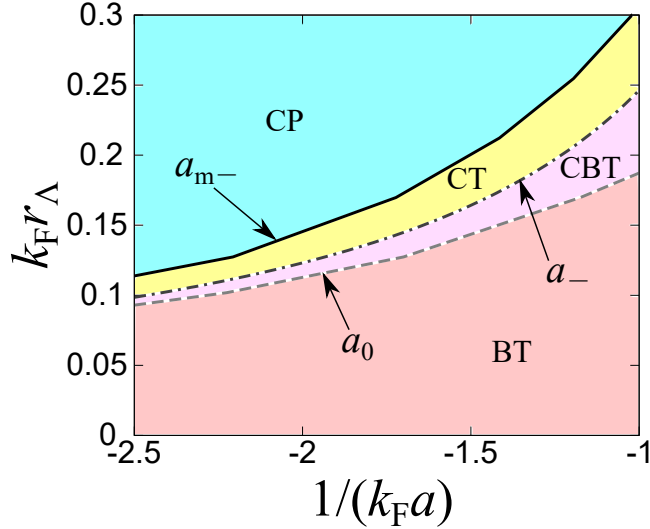}
    \caption{Ground-state phase diagram in the plane of the range parameter $k_{\rm F}r_\Lambda$ and the inverse scattering length $1/(k_{\rm F}a)$.
    The characteristic couplings $1/(k_{\rm F}a_{{\rm m}-})$, $1/(k_{\rm F}a_{-})$, and $1/(k_{\rm F}a_0)$ that distinguish between CP, CT, CBT, and TB are plotted.}
    \label{fig:3}
\end{figure}

We summarize the ground-state phase diagram in the space of $k_{\rm F}r_\Lambda$ and $1/(k_{\rm F}a)$ in Fig.~\ref{fig:3}.
While the Cooper pairing dominates the sufficiently long range regime,
the system undergoes the formation of bound trimers in the sufficiently short range regime.
In between, one can find non-trivial Cooper triples and correlated bound trimers to occur.
While it is similar to the phase diagram for the Lifshitz transition at $a=a_{-}$ from the trimer Fermi liquid to the superfluid state at negative scattering lengths within the mean-field and in-vacuum three-body analyses~\cite{Nishida2012},
our results imply that competing fluctuations associated with the pair-to-triple transition give a rich physics on this system. 
Such a possibility has also been predicted in a diagrammatic approach at finite temperature~\cite{Tajima2019}.
Also, the crossover from the Cooper triple to the tightly-bound trimer state is analogous to the BCS-BEC crossover near the Feshbach resonance where Cooper pairs continuously change to tightly-bound molecules.

\begin{figure}[t]
    \centering
    \includegraphics[width=8.5cm]{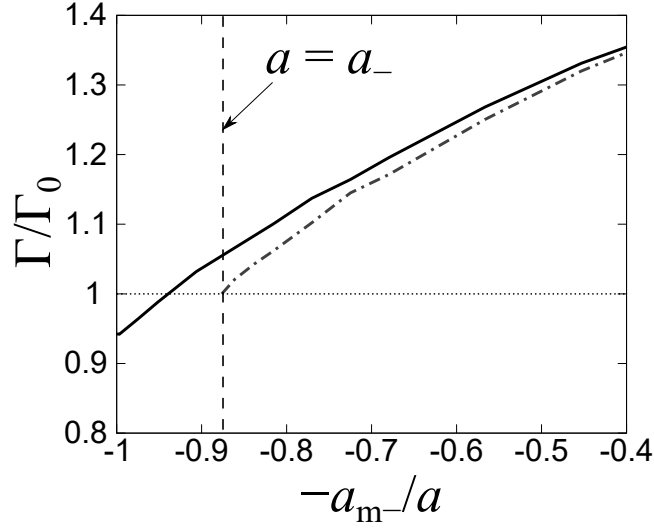}
    \caption{Three-body loss rate $\Gamma$ in the crossover from the Cooper triple to the bound trimer state (solid curve) at $k_{\rm F}r_\Lambda=0.127$.
    The dashed-dotted curve shows the result in the absence of the Fermi surface. The calculations are stopped where the three-body bound-state solution disappears.  We normalize $\Gamma$ with $\Gamma_0$ which is the three-body loss rate at $a=a_{-}$ in the vacuum case.}
    \label{fig:4}
\end{figure}

For experimental observations, we expect that in the presence of the Fermi surface, the three-body loss becomes large already at $a=a_{{\rm m}-}$.  This is in contrast to the usual expectation that it does at $a=a_{-}$.
To see this, we consider the non-Hermitian three-body term $W$~\cite{Kirk} for the three-body loss given by
\begin{align}
    W=-i\gamma\sum_{\bm{k},\bm{k}',\bm{p},\bm{p}',\bm{q}}c_{\bm{k},1}^\dag c_{\bm{p},2}^\dag c_{\bm{q}-\bm{k}-\bm{p},3}^\dag c_{\bm{q}-\bm{k}'-\bm{p}',3} c_{\bm{p}',2} c_{\bm{k}',1},
\end{align}
where $\gamma$ is a real-valued strength for the decay.
The norm of the variational wave function $\langle \Psi_{\rm CT}|\Psi_{\rm CT}\rangle$ is not conserved in the presence of $W$.
On the basis of the small perturbation with respect to $W$,
one can define the three-body loss rate $\Gamma$~\cite{Kirk} as
\begin{align}
    \Gamma&=2i\langle \Psi_{\rm CT}|W|\Psi_{\rm CT}\rangle
    =2\gamma\sum_{\bm{p},\bm{k}}\mathcal{A}^*(\bm{p})\mathcal{A}(\bm{k}).
\end{align}
We have calculated $\Gamma$ as a function of $a$.
Figure~\ref{fig:4} shows the result at $k_{\rm F}r_\Lambda=0.127$ as normalized by the in-vacuum result $\Gamma_0$ at $a=a_{-}$.
We find from this figure that the three-body loss is amplified in the Cooper triple regime and gradually approaches the in-vacuum counterpart (dashed-dotted curve in Fig.~\ref{fig:4}) as $|a|$ increases.
In this sense, the amplification of the three-body loss in the presence of the Fermi surface compared to the in-vacuum case could be a signature of the Cooper triple formation.
Although such a shift originating from the Fermi sea has already been predicted in a different situation where two of three hyperfine states have Fermi seas~\cite{Nygaard,Sun2019,Sanayei2020},
our result gives a physical interpretation based on the Cooper triple state. 
We note that the calculation of $\Gamma$ is stopped at $a=a_{{\rm m}-}$ ($a=a_-$ for the in-vacuum case) where the in-medium three-body bound state disappears. 
In general, $\Gamma$ is finite even in the absence of the three-body bound state but it is expected to be small as observed in the experiments to probe the Efimov effects~\cite{NaidonEndo}.
While we consider symmetric two-body interactions,
such a situation can be realized in three-component $^6$Li Fermi gases at a high magnetic field~\cite{Ohara}.

\section{Summary}
\label{sec5}
To summarize, we have investigated the transition from the Cooper pair to the Cooper triple state and the three-body crossover from the Cooper triple to the tightly-bound trimer state near the triatomic resonance in a three-component Fermi gas with two-body attractive contact interactions.
Using the variational ansatz for two- and three-body states on top of the Fermi sea,
we calculate the in-medium pair and trimer binding energies at arbitrary scattering lengths and range parameters.
\par
From the analysis of the three-body pole in the in-medium STM equation, we determine the four regimes dominated by Cooper pairs, Cooper triples, correlated bound trimers, and bound trimers, respectively. 
The competition between Cooper pairs and triples occurs in the weak-coupling regime.
Near the triatomic resonance where a three-body bound state appears in vacuum, Cooper triples undergo the crossover to tightly bound trimers with increasing inverse scattering length.
This crossover is analogous to the BCS-BEC crossover near the Feshbach resonance in a two-component Fermi gas.
Such transition and crossover also occur when the range parameter decreases. 
We present the ground-state phase diagram of this system in the space of the the inverse scattering length and the two-body range parameter associated with the momentum cutoff.
\par
For experimental observations, we predict the shift of the threshold from $a_{-}$ to $a_{{\rm m}-}$ for the trimer formation due to the Fermi surface effect. 
Cooper triples exist in the region of $a_{{\rm m}-}<a<a_{-}$.
Moreover, we show that the three-body loss rate is amplified by the formation of Cooper triples and correlated bound trimers.
Thus, the three-body loss at low temperatures and moderate densities, where quantum degeneracy is kept, can be a useful indication for the emergence of such non-trivial states in cold atom experiments.  
\par
In this study, we do not consider the three-body interaction, which may change the position of the in-medium triatomic resonance.
Also, full time evolution of in-medium clusters for a quantitative comparison with experiments should be addressed in the future.
Both of them can be implemented in the present variational approach.
Moreover, it is important to investigate the finite temperature and density phase diagram with more sophisticated many-body approaches.
Cooper triples in mass- and spin-imbalanced systems and a four-body counterpart in four-component Fermi gases are also interesting future problems.
Another interesting direction is to investigate the formation of Cooper triples in quark matter and its competition with color superconductivity.
This will be presented elsewhere.

\acknowledgements
H.\ T. thanks P. Naidon, S. Endo, M. Horikoshi, Y. Nishida, T. Hatsuda, and S. Akagami for useful discussions in the initial stage of this study and 
the members of H. Liang's group in The University of Tokyo and of theoretical physics group in Kochi University for helpful comments.
H.\ T. and K.\ I. were supported by Grants-in-Aid for Scientific Research from JSPS (No.\ 18H05406).
S.\ T.\ was supported by the RIKEN Special Postdoctoral Researchers Program.
T.\ M.\ D. was supported by Grant-in-Aid for Early-Career Scientists (No. 20K14480).
K.\ I. was supported by Grants-in-Aid for Scientific Research from JSPS (18H01211).
This work was partly supported by RIKEN iTHEMS Program.

\end{document}